\documentclass[twocolumn,aps,prl,amsmath,floatfix]{revtex4}
\usepackage{graphicx}
\begin{document}

\title{Novel Mott Transitions in the Nonisotropic Two-Band Hubbard Model}
\author{A. Liebsch} 
\affiliation{Institut f\"ur Festk\"orperforschung, 
             Forschungszentrum J\"ulich, 
             52425 J\"ulich, Germany}
\begin{abstract}
The Mott transition in a two-band Hubbard model involving subbands of 
different widths is studied as a function of temperature using dynamical 
mean field theory combined with exact diagonalization. The phase diagram 
is shown to exhibit two successive first-order transitions if the full
Hund's rule coupling is included. In the absence of spin-flip and 
pair-exchange terms the lower transition remains first-order while the
upper becomes continuous.
\end{abstract}
\maketitle

The nature of the metal insulator transition in materials involving
subbands of different widths has been intensively debated during the
recent years \cite{anisimov,epl,prl,koga1,prb,koga2,koga3,ferrero,%
medici,arita,song,knecht}. 
This issue is relevant for the understanding of the effect of strong
local Coulomb interactions in systems such as Ca$_{2-x}$Sr$_x$RuO$_4$. 
In this layer perovskite the partially filled Ru $t_{2g}$ bands for $x=2$ 
consist of wide $d_{xy}$ and narrow $d_{xz,yz}$ bands \cite{oguchi}. 
As pointed out in \cite{prl2000}, the onsite Coulomb energy lies between 
the subband widths: $W_{xz,yz}<U<W_{xy}$. The usual criterion with the 
parameter $U/W$ as a measure of the importance of correlations must then 
be generalized. The pure Sr compound is superconducting below 1.5 K 
\cite{maeno}. 
Iso-electronic replacement of Sr by Ca leads to an effective band 
narrowing due to octahedral distortions \cite{fang} 
and a metal insulator transition \cite{nakatsuji}. 
As a consequence of non-cubic crystal fields many other transition metal 
oxides also involve non-equivalent partially occupied subbands.   

A key question in these materials is whether the wide and narrow subbands
exhibit separate Mott transitions or whether single-particle hybridization
 and inter-orbital Coulomb interactions ensure a single transition for all 
bands simultaneously. This issue was studied initially by Anisimov et 
al.~\cite{anisimov} and \mbox{Liebsch} \cite{epl} for Ca$_{2-x}$Sr$_x$RuO$_4$ 
using simplified band structure models which did not yet include the full 
complexity due to Ca-induced octahedral distortions. Correlations were
treated in the dynamical mean field theory \cite{dmft} (DMFT) combined 
with the non-crossing approximation (NCA) and Quantum Monte Carlo (QMC) 
method, respectively. Since the NCA calculations (at $T=0$) neglected 
interorbital Coulomb interactions the results showed separate, 
`orbital-selective' Mott transitions for the narrow and wide subbands 
\cite{anisimov}. In contrast, the QMC calculations (at $T=0.125$~eV) 
included interorbital Coulomb interactions and suggested a common transition 
for all $t_{2g}$ bands \cite{epl}. 

Recent theoretical studies of the Mott transition in a paramagnetic 
two-band model system have led to apparent contradictions
\cite{prl,koga1,prb,koga2,koga3,ferrero,medici,arita}.  
Including the full Hund's rule coupling Koga {\it et al.}~\cite{koga1} 
found orbital-selective metal insulator transitions at $T=0$. 
Neglecting spin-flip and pair-exchange terms Liebsch \cite{prb} obtained 
at $T>0$ a single first-order transition, followed by a 
bad-metallic/non-Fermi-liquid (NFL) phase. 
As we argue below, the various results and conclusions are 
consistent provided that the choice of crucial parameters such as 
temperature and Hund's rule coupling is properly taken into account.

In the present work we combine finite temperature DMFT with exact 
diagonalization \cite{ed} (ED) to determine the $T/U$ phase diagram 
of a two-band system consisting of non-hybridizing, half-filled subbands 
with semi-circular density of states of width $W_1=2$~eV and $W_2=4$~eV. 
The subbands interact 
via intra- and interorbital Coulomb matrix elements $U$ and $U'=U-2J$, 
where $J$ is the Hund's rule exchange integral. In contrast to the 
multi-band QMC approach, which includes only Ising-like exchange terms to 
avoid sign problems at low temperatures \cite{held}, ED permits also the
consideration of spin-flip and pair-exchange interactions. To be specific 
we take $J=U/4$ which is approximately satisfied in several transition 
metal oxides. 

The main result of this work is that the $T/U$ phase diagram in the 
presence of the full Hund's rule exchange exhibits 
{\it two successive first-order phase transitions}, with separate  
hysteresis loops and coexistence regions. The intermediate region 
corresponds to the $T>0$ analog of the orbital-selective Mott 
(OSM) phase obtained at \mbox{$T=0$} in \cite{koga1}. On the other hand, 
if spin-flip and pair-exchange terms are omitted, we find a 
{\it single first-order transition succeeded by a non-Fermi-liquid 
phase}, in agreement with previous QMC results \cite{prb}. 
Both trends are consistent with those obtained by several groups 
\cite{koga1,koga2,ferrero,medici,arita} for the same two-band model 
at $T=0$.

The ED/DMFT results are derived from a two-band generalization of the 
approach employed for single-bands \cite{dmft,ed,gracias}. Since at $T>0$ 
all states of the impurity Hamiltonian are used in the construction 
of the subband Green's functions $G_i(i\omega_n)$, two bath 
levels per impurity level are taken into account ($n_s=6$ per spin). 
To check the accuracy of this approximation we have evaluated the $T/U$ 
phase diagram of a single band for $n_s=3,\dots,6$. As shown in Fig.~1,
the stability boundaries $U_{c1}(T)$ and $U_{c2}(T)$ for $n_s=3$ are 
slightly too low. Nevertheless, the overall shape of the phase diagram 
agrees qualitatively with the converged results for $n_s=6$ \cite{tong}. 
Thus, we are confident that in the two-band case $n_s=6$ also yields a 
reasonable picture of the phase diagram. Preliminary results for $n_s=8$ 
will also be presented. 

\begin{figure}[t!]
  \begin{center}
  \includegraphics[width=4.5cm,height=8cm,angle=-90]{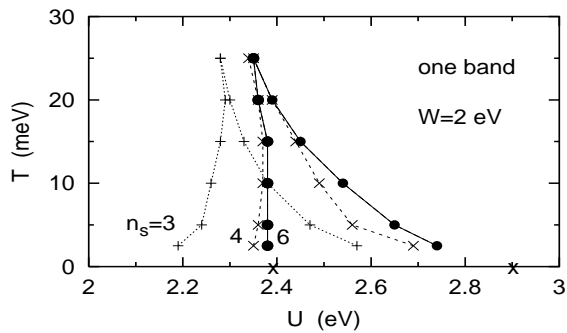}
  \end{center}
  \vskip-2mm
\caption{
Phase diagram for one-band Hubbard model, calculated within ED/DMFT
for $n_s=3,4,6$. Symbols (x): $T=0$ transitions obtained in \cite{bulla}.
}\end{figure}

To analyze the metal/insulator transition we study the quasiparticle weights 
$Z_i = 1/[1-d{\rm Re}\,\Sigma_i(\omega)/d\omega\vert_{\omega=0}]$, 
which in the metallic range can be represented as
$Z_i \approx 1/[1- {\rm Im}\,\Sigma_i(i\omega_0)/ \omega_0]$, where 
$\Sigma_i(i\omega_0)$ is the subband self-energy at the first Matsubara 
frequency. Fig.~2(a) shows $Z_i(U)$ for $J'=J=U/4$, where $J'$ 
denotes spin-flip and pair-exchange terms \cite{fit}. Two critical regions 
can be identified, each with hysteresis loops characteristic of first-order 
phase transitions. The coexistence areas are $U^<_{c1}(T) < U < U^<_{c2}(T)$ 
near 2.0~eV and $U^>_{c1}(T) < U < U^>_{c2}(T)$ near 3.0~eV, where 
$U^<_{cn}(T)$ and $U^>_{cn}(T)$ are the stability boundaries obtained for 
increasing ($n=2$) and decreasing ($n=1$) $U$.
Let us denote the true critical energies of these transitions
as $U^<_c(T)$ and $U^>_c(T)$. Below $U^<_c(T)$ both bands are metallic 
while above $U^>_c(T)$ both are insulating. At the lower transition 
both bands undergo first-order transitions -- but in fundamentally 
different ways: $Z_1(U)$ becomes very small while $Z_2(U)$ drops to a 
finite value. The narrow band therefore undergoes a `complete' 
metal/insulator transition, whereas the wide band exhibits an `incomplete' 
transition to a new, considerably more correlated phase. This band becomes 
fully insulating near 3.0~eV, where it exhibits a weak second hysteresis loop. 
To our knowledge this is the first time that {\it sequential first-order 
transitions} are identified in the $T/U$ phase diagram of a Hubbard model 
involving non-equivalent bands \cite{kawakami}.
If such a material could be encountered experimentally, the conductivity 
as a function of pressure would show two consecutive jumps. 

Fig.~2(b) shows $Z_i(U)$ for $J'=0$, $J=U/4$, i.e., in the absence of 
spin-flip and pair-exchange terms.  The results are similar to those in 
Fig.~2(a), with the important exception that the wide band above the lower
transition is even more correlated and the upper transition is now continuous 
at smaller $U$ \cite{continuous}. 
Remarkably, the lower transition remains first-order for both subbands. 
In this case the conductivity shows a jump at $U^<_c(T)$ but a change of slope 
at $U^>_c(T)$. The results in Fig.~2(b) confirm the picture obtained previously 
within the QMC for $T>0$, $J'=0$ which showed the existence of a single 
first-order transition followed by a mixed insulating/bad-metallic phase 
\cite{prb,ipt,bluemer1}. They also demonstrate that the two-band ED/DMFT 
for $n_s=6$ is qualitatively accurate \cite{hyst}. 

\begin{figure}[t!]
  \begin{center}
  \includegraphics[width=8cm,height=8cm,angle=-90]{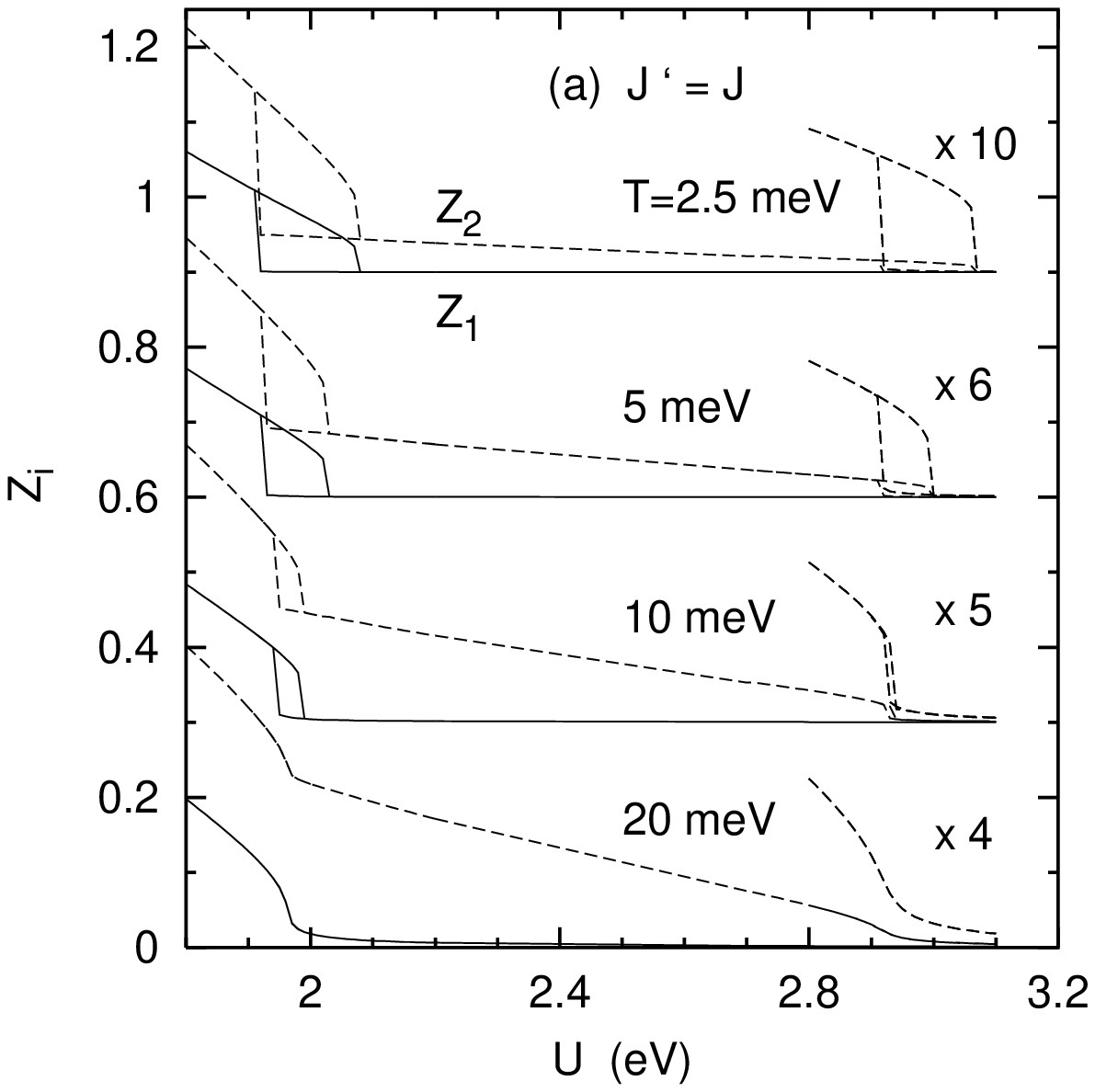}
  \includegraphics[width=8cm,height=8cm,angle=-90]{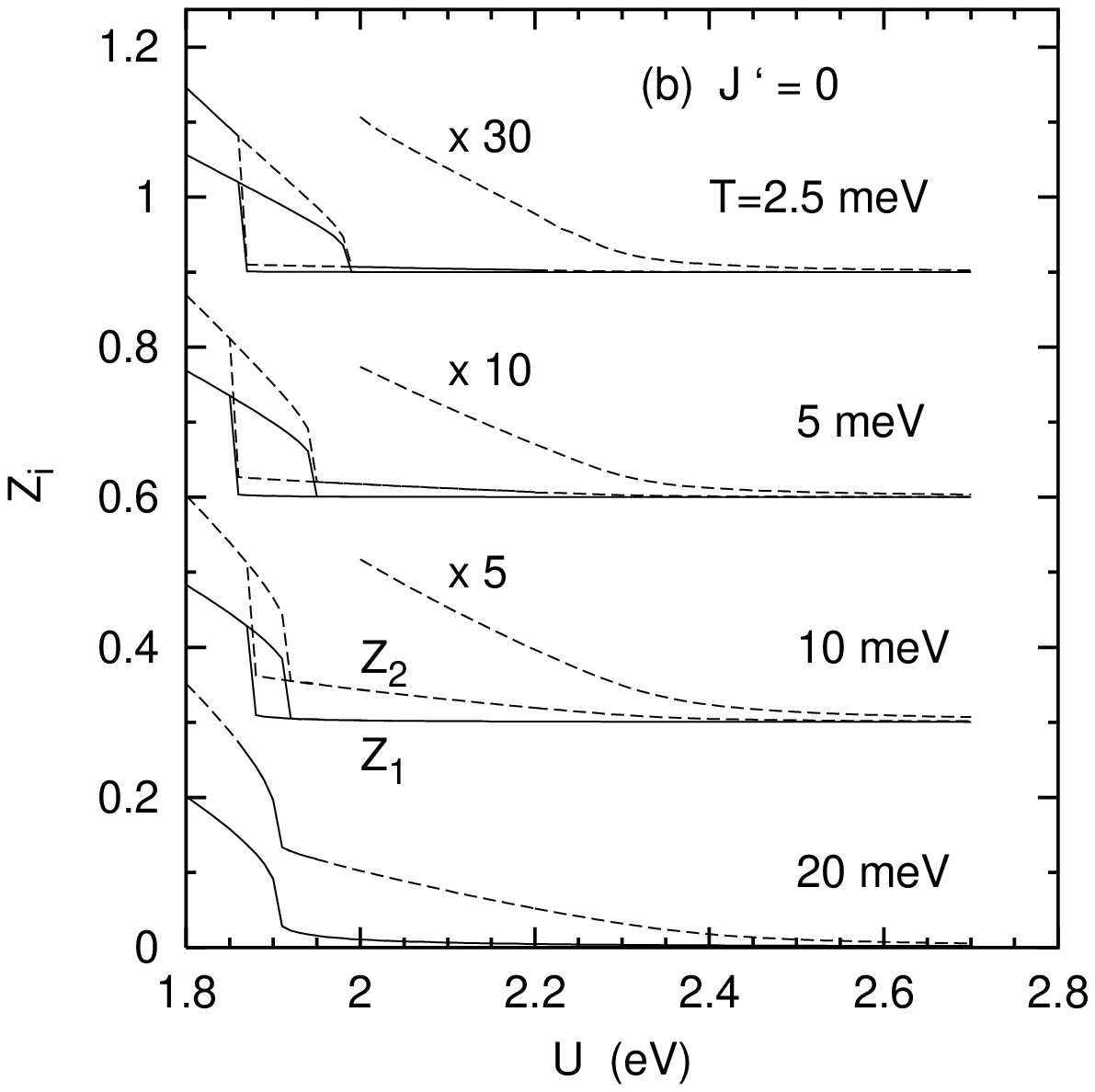}
  \end{center}
  \vskip-2mm
\caption{
$Z_i(U)$ of nonisotropic two-band Hubbard model, calculated within ED/DMFT. 
(a) $J'=J=U/4$, (b) $J'=0$, $J=U/4$. Solid (dashed) curves: narrow (wide)
band. Results for different temperatures are displaced vertically by $0.3$.
}\end{figure}

The phase diagrams deduced from the ED results for $T\ge2.5$~meV are shown 
in Fig.~3. For $J'=J$ as well as $J'=0$ the transition at $U^<_c(T)$ is 
first-order for both subbands. The subsequent transition of the wide band at 
$U^>_c(T)$ is first-order for $J'=J$ but continuous for $J'=0$.  
At $U^<_c(T)$ the metal/insulator transition is complete only for the narrow 
band. The wide band first undergoes a transition to a more strongly correlated 
phase and becomes truly insulating at the second transition at $U^>_c(T)$. 
The overall shape of the phase diagram for $J'=J$ agrees with the one 
recently obtained by Inaba {\it et al.} \cite{kawakami}. 

\begin{figure}[t!]
  \begin{center}
  \includegraphics[width=5cm,height=8cm,angle=-90]{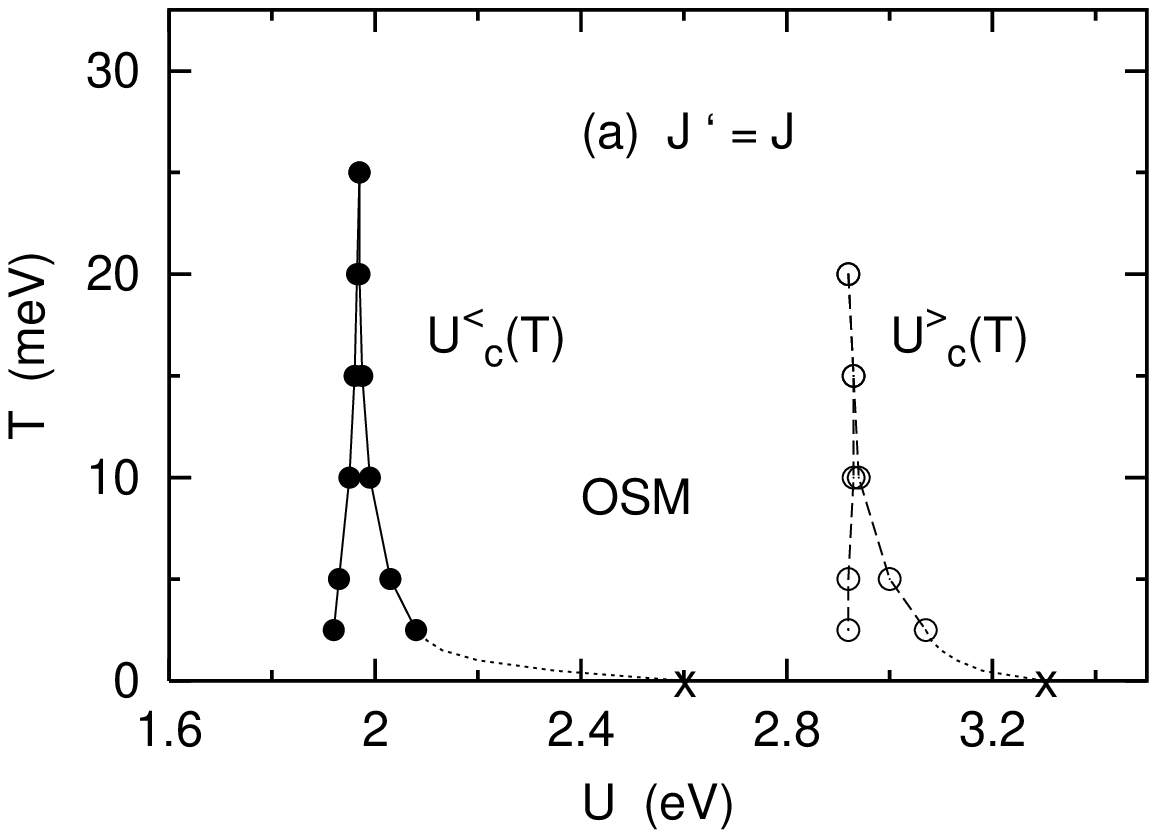}
  \includegraphics[width=5cm,height=8cm,angle=-90]{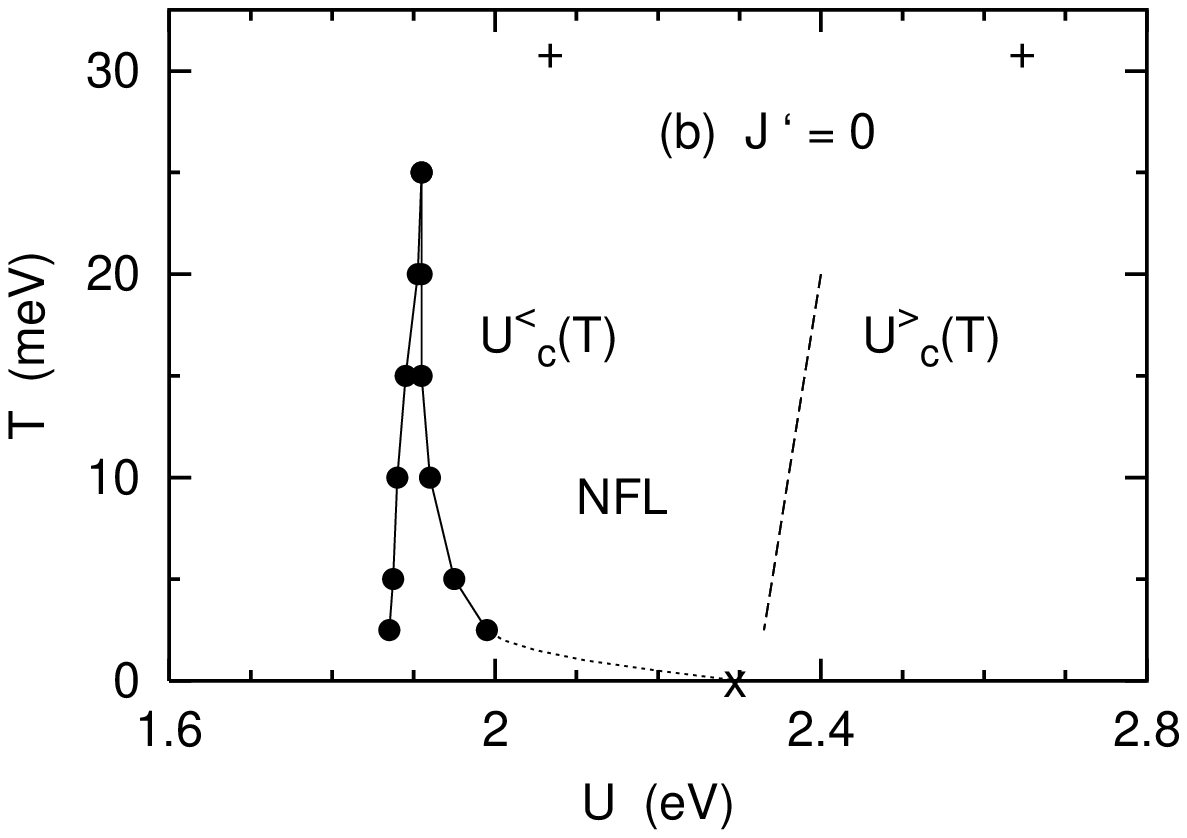}
  \end{center}
  \vskip-2mm
\caption{
Phase diagram for nonisotropic two-band Hubbard model, calculated within 
ED/DMFT. (a) $J'=J=U/4$, (b) $J'=0$, $J=U/4$. Solid dots in (a), (b): 
stability boundaries of both subbands near lower first-order transition.
Open dots in (a): stability boundaries of wide band near upper first-order 
transition. Dashed line in (b): approximate location of continuous transition 
of wide band. Symbols (x): $T=0$ transitions obtained in \cite{koga1,koga2,arita};
(+): transitions at $T=31$~meV  obtained in \cite{prb,knecht}.    
Lines are guides to the eye.
}\end{figure}

Fig.~3 also shows the critical Coulomb energies obtained at $T=0$ 
\cite{koga1,koga2,arita} and $T=31$~meV \cite{prb,knecht}. As illustrated 
in Fig.~1 for the one-band case, the ED scheme with small $n_s$ underestimates 
the critical Coulomb energies by about 0.1 to 0.2~eV. Preliminary results 
for $n_s=8$ indicate similar shifts in the two-band case. Taking these 
displacements into account the ED results shown in Fig.~3 are in excellent 
correspondence with those obtained at $T=0$ and $T=31$~meV.
Evidently the conflicting conclusions reached in 
\cite{anisimov,epl,prl,koga1,prb,koga2,ferrero,medici,arita} concerning 
the nature of the Mott transition in multi-band systems were caused by 
different behaviors obtained for $T=0$ vs. $T>0$ and $J'=J$ vs. $J'=0$. 
Accounting for these different parameter choices, the DMFT treatments 
are consistent. 

Quasiparticle spectra derived within the QMC/DMFT for $J'=0$ \cite{prb} 
showed that in the intermediate phase the self-energy of the wide band  
at small $\omega_n$ deviates significantly from metallic $\sim\omega_n$ 
behavior \cite{biermann}. Accordingly, the quasi-particle spectra show 
a pseudogap which for larger $U$ gets more pronounced \cite{bad}, indicating 
progressive non-Fermi-liquid properties. A true gap opens at $U^>_c(T)$. 
Thus, $Z_2(U)>0$ 
in the region $U^<_c(T)<U<U^>_c(T)$ does not imply existence of quasiparticles. 
As indicated in Fig.~3(b) the intermediate `orbital-selective Mott' phase in 
the absence of spin-flip and pair-exchange is in fact a mixed insulating/NFL 
phase. The same trend is found using the $T>0$ ED/DMFT \cite{future}. 

\begin{figure}[t!]
  \begin{center}
  \includegraphics[width=5cm,height=4cm,angle=-90]{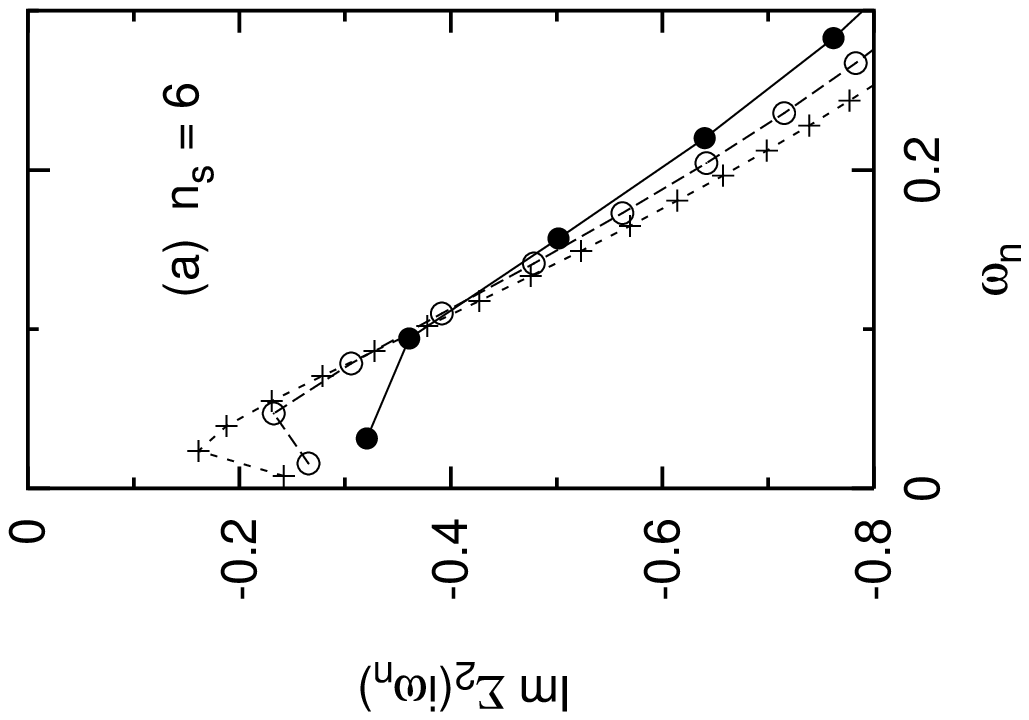}\hskip-3mm
  \includegraphics[width=5cm,height=4cm,angle=-90]{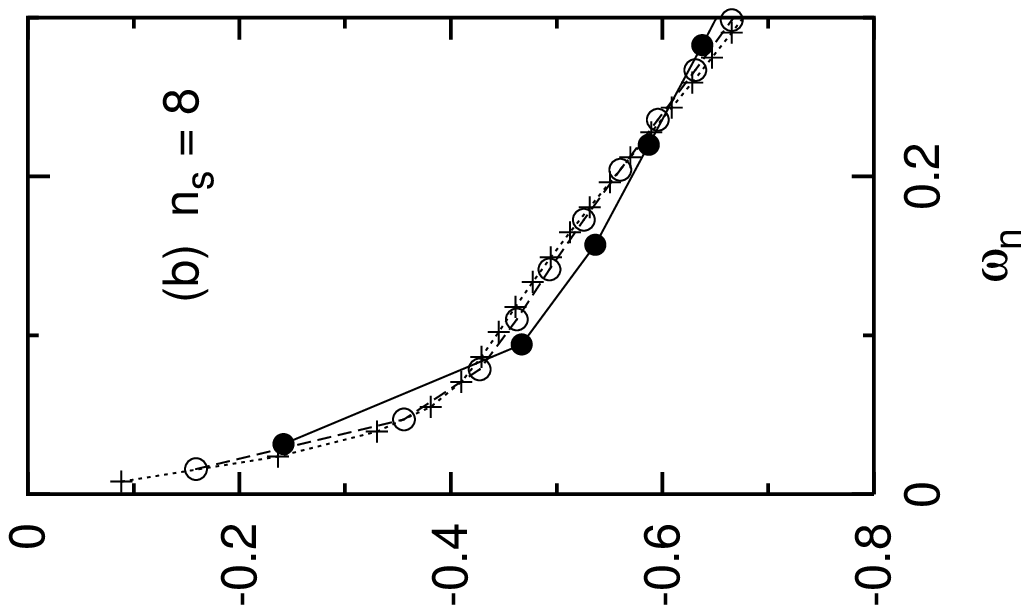}
  \end{center}
  \vskip-2mm
\caption{
Self-energy of wide band for $J'=J$ in intermediate phase at $U=2.4$~eV. 
(a) $n_s=6$; (b) $n_s=8$. Solid dots: $T=10$~meV; empty dots: $T=5$~meV; 
(+): $T=2.5$~meV. 
}\end{figure}

For $J'=J$, the low-frequency analysis of $\Sigma_2(i\omega_n)$ is more
intricate. As shown in Fig.~4(a) for $n_s=6$, $\Sigma_2(i\omega_n)$ 
reveals deviations from metallic $\sim\omega_n$ variation, giving rise to 
small pseudogaps in the quasiparticle spectra. This behavior is incompatible 
with Im\,$\Sigma_2(\omega)\sim\omega^2$ and suggests that, as for $J'=0$, in 
the intermediate phase the wide band at $T>0$ violates Fermi-liquid theory. 
The extension of the present ED approach to $n_s=8$ indicates, however, that 
the additional bath levels are important for the low-frequency variation of 
$\Sigma_2(i\omega_n)$. As can be seen in Fig.~4(b), the deviations are absent 
and the $\sim\omega_n$ variation is consistent with Fermi-liquid behavior.
In fact, the shoulder near $\omega_0=0.06$ suggests that Fermi-liquid 
properties persist up to about $T=\omega_0/\pi\approx 20$~meV. Thus, 
the OSM phase in Fig.~3(a) is the $T>0$ analog of the orbital-selective 
Mott phase identified first by Koga {\it el al.}~\cite{koga1} at $T=0$.
A more complete discussion of the results for $n_s=8$ will be given elsewhere 
\cite{future}. Because of finite size limitations of the present ED/DMFT 
scheme associated with the small number of bath levels, a precise 
determination of low-temperature properties is not possible. Nevertheless,      
an approximate extrapolation of $\Sigma_2(i\omega_n)$ indicates that the 
$T\rightarrow0$ limit for $J'=J$ satisfies Fermi-liquid criteria, in 
agreement with previous $T=0$ studies \cite{koga1,koga2,ferrero,medici,arita}.

In view of the importance of spin-flip and pair-exchange terms for 
the Mott transition in multi-band materials \cite{pruschke} it is 
desirable to investigate the $T/U$ phase diagram by methods 
which permit adequate treatment of the complete Hund's rule matrix, 
for instance, a two-band extension of $T>0$ numerical renormalization
group studies \cite{costi}; see also \cite{kawakami}.   
Because of sign problems, recent QMC extensions including spin-flip 
and pair-exchange terms are limited to $T\ge 1/6$~eV$\ \gg T_c$ 
\cite{koga3} or $T=0$ \cite{arita}. Also, other recent works 
\cite{ferrero,medici} employing a variety of quantum impurity 
methods deal so far mainly with $T=0$ and do not yet allow the 
identification of the multi-band Mott transition at general $T,U$ 
values. 

In summary, the $T/U$ phase diagram of the Hubbard model involving 
half-filled, non-equivalent subbands is shown to be remarkably rich.
The competing kinetic energy scales, coupled via Coulomb and exchange 
energies, give rise to {\it sequential first-order phase transitions}. 
The lower transition separates a purely metallic phase from a mixed
phase with insulating narrow and strongly-correlated wide subbands. 
The wide band becomes insulating at the second first-order transition. 
Omission of spin-flip and pair-exchange terms enhances the correlations 
in the wide band in the intermediate phase so that it no 
longer satisfies Fermi-liquid criteria, and modifies the upper phase 
transition from first-order to continuous.  

For the analysis of experimental data of materials such as 
Ca$_{2-x}$Sr$_x$RuO$_4$ it is necessary to account also for hybridization 
between orbitals. Preliminary studies of this effect within two-band 
models for $T\gg T_c$ \cite{koga3} and $T=0$ \cite{medici,song} suggest 
significant changes. Moreover, spatial fluctuations \cite{kotliar} and 
deviations from half-filling might play a decisive role close to the 
Mott transition. More work is needed to investigate whether both 
first-order transitions persist in the presence of these effects, or 
whether the weak first-order behavior of the upper transition disappears 
and only the dominant lower transition survives as the common first-order 
Mott transition for all bands. 

I like to thank Th. Costi, K. Held, N. Kawakami, and E. Koch for valuable 
discussions.

Email: a.liebsch@fz-juelich.de


\begin{thebibliography}{99}
\bibitem{fit}
   More accurate results at low $\omega_n$ and $T$ are obtained by minimizing 
   differences between impurity and Anderson model Green's functions via 
   $\sum_n \vert\Delta G_i(i\omega_n)\vert^2$ rather than 
   $\sum_n \vert\Delta G_i(i\omega_n)\vert$. I thank E. Koch for this suggestion.
\bibitem{continuous}
   The term `continuous' indicates here that there is no jump; it may
   also signify crossover behavior.
\bibitem{hyst}
   The QMC hysteresis region is slightly wider than in ED.
   This could be due to slow convergence of QMC at $T=0.02$~eV 
   and too small $n_s$ in ED.
\end{thebibliography}
\end{document}